# Actions francophones autour des normes e-learning à l'ISO

## Pour un accès multilingue et multiculturel égalitaire à l'éducation

**Mokhtar Ben Henda\* — Henri Hudrisier\*\***

*\* ISIC, Université de Bordeaux*
*33607 Pessac, FRANCE*

*Mokhtar.benhenda@u-bordeaux3.fr*
*\*\* Université Paris 8, MSH Paris Nord,*
*93000 Saint Denis, FRANCE*

*henri.hudrisier@wanadoo.fr*

RESUME : *L'avenir de l'*e-Learning *est en train de se construire dans les instances de normalisation et de standardisation des TIC à échelle mondiale, tout particulièrement au sein du sous comité 36 de l'ISO, chargé de normaliser les technologies éducatives. Les auteurs de cette communication, délégués officiels de l'AUF (Agence Universitaire de la Francophonie) auprès de cette structure, mettent en évidence l'importance stratégique du suivi participatif à ces actions de normalisation pour préserver la diversité culturelle, linguistique et une égalité d'accès à l'éducation pour tous[1].*

ABSTRACT: *The future of* e-Learning *is on the way to be constructed within ICT standardization international instances. The sub-committee 36 of ISO, which is responsible for standardizing educational technologies, is certainly the most prominent of all. The authors of this paper, who are official delegates of the Agency of French Speaking Universities (AUF) with this structure, highlight the strategic importance of active monitoring of* e-Learning *standards for preserving cultural diversity, linguistic and equal access to education for all.*

---

[1] *Ce texte est avant tout une réflexion sur la question de la gouvernance du e-learning. Il présente et analyse dans les instances de normalisation, les positions de la délégation française dans le contexte global des TIC puis sous l'angle spécifique au e-learning. Le texte et avance aussi des propositions en matière de diversité linguistique. il pourra servir de point d'ancrage à un débat autour de ces thèmes (Note du conseil scientifique).*

**1. Introduction**

Les TICE (Technologies de l'Information et de la Communication pour l'Education) sont depuis longtemps l'objet d'une standardisation de méthodes, de contenus et d'outils. Des structures pionnières telles comme AICC[2], IMS[3], ou ARIADNE[4], ont longuement œuvré pour instancier leurs propres conceptions des stratégies et démarches de gestion et d'organisation des TICE. Pour ces pionniers, notamment l'aéronautique (AICC), et la formation militaire (ADL), l'e-Learning était un enjeu de réussite primordial. Ils ont contribué ainsi à la genèse d'un éventail de recommandations, spécifications, standards et normes ayant trait à gérer un ou plusieurs aspects de l'activité d'enseignement, de formation ou d'apprentissage. Ce sont encore ces mêmes pionniers qui en 1999, ont saisi l'ISO[5] en suscitant la création du sous-comité 36[6] qui a été créé en mars 2000 à Londres, pour mettre en chantier la normalisation des TICE sur la base d'une synthèse des standards et des spécifications en vigueur. Mais l'objectif est aussi de proposer la production d'une famille de normes e-Learning qui saurait répondre aux attentes de tous, dans un souci d'équité d'accès, de partage et de mutualisation des acquis éducatifs. Ce dernier enjeu, est on le perçoit beaucoup plus prospectif. Il ne s'agit pas seulement de définir un dénominateur commun des standards d'e-Learning existants[7], mais de projeter vers l'avenir les spécifications normatives d'un cadre de développement recherche qui devient un socle commun et consensuel pour les acteurs des TICE. Il

---

[2] Aviation Industry CT Committee. http://aicc.org/ , [visité le 4 avril 2007]
[3] Instructional Management System de Global Learning Consortium. http://www.imsglobal.org/ [visité le 4 avril 2007
[4] Association of Remote Instructional Authoring and Distribution Networks for Europe. http://www.ariadne-eu.org [visité le 4 avril 2007]
[5] L'ISO est un réseau d'instituts nationaux de normalisation de 157 pays, selon le principe d'un membre par pays, dont le Secrétariat central, situé à Genève, assure la coordination d'ensemble.  ISO (originellement *International Organization for Standardization*), ce sigle se décline aujourd'hui en anglais : *International Organization for Standardization* et en français Organisation Internationale de Normalisation. C'est donc un acronyme volontairement international (en référence à la racine grecque *ISO*, égal)
[6] L'ISO/IEC JTC1 SC36 (en forme courte le SC36). Le sigle se développe ainsi :

ISO/IEC (*International Electrotechnical Commission* ; JTC1 (*Joint Technical Committee 1*, un Comité technique commun à l'ISO et à l'IEC) ; SC36 (Sous-comité n° 36 du JTC1).

[7] Comme cela se fait souvent dans des chantiers de normalisation sur des secteurs à cycle d'innovation beaucoup plus lent.

s'agit pour faire image de mettre en place non plus seulement les contraintes architecturales[8], mais les contraintes normatives d'un urbanisme des TICE.

Cependant, toujours au tournant des années 2000, l'intérêt pour l'e-Learning[9], avait considérablement évolué sous l'impulsion de la mondialisation et des réseaux informatiques ouverts et distribués. On assistait déjà à l'émergence d'une mosaïque de solutions et de démarches pédagogiques ce qui contribuait bien sûr à rendre plus problématique les questions de compatibilité et d'interopérabilité.

Or, aujourd'hui, le chantier normatif du SC36 pour l'e-Learning est encore en quête de résultats probants nécessitant un consensus universel et une représentativité équilibrée de toutes les populations mondiales concernées par les effets que la future norme e-Learning engendrerait sur les politiques et les stratégies éducatives dans le monde. C'est justement sur cette question de représentativité mondiale dans une structure aussi marquante sur le plan stratégique de l'e-Learning, que nous orienterons le contenu de cet article pour justifier l'action francophone au sein de l'ISO afin de défendre les intérêts culturels et linguistiques des protagonistes absents[10] de ce chantier stratégique.

Nous soulèverons d'abord un certain nombre de points autours des questions clefs suivantes : comment la normalisation crée un cadre mondialement concerté du développement et de la recherche des TIC (et par voie de conséquence des TICE) ? Quel rôle cela joue dans la convergence multimédia des TIC et de leur interopérabilité grandissante, la convergence notamment du e-Learning avec le *m-Learning* (*m* pour mobile) et le *t-Learning* (*t* pour télévision numérique) ? Comment les processus de description normalisée des ressources d'enseignement sont-ils une des conditions clefs de leur échange et de leur réutilisabilité tant interdisciplinaire, qu'internationale, interculturelle et/ou inter linguistique ?

**2. Normalisation en e-Learning : une quête d'universalité**

Nous partons du postulat qu'une participation active aux instances internationales de normalisation contribue théoriquement à aménager un cadre d'échange des ressources, mais permet aussi que puissent se développer des terminaux d'accès ou des réseaux, réellement respectueux des diversités linguistiques, géopolitiques, disciplinaires et institutionnelles. Défendre dans ces commissions le droit à la diversité culturelle, linguistique, économique permet aussi de prévenir l'aggravation de la fracture numérique Nord-Sud. C'est aussi en étant vigilants à toutes les étapes de production des consensus qui fondent les normes qu'on peut préserver maints autres types de possibilité d'accès au savoir tout en préservant les libertés individuelles et les biens communs universels de transmission de la connaissance.

---

[8] Au sens où on l'entend aussi dans la conception des TIC.
[9] Nous ferons du concept e-Learning dans ce papier, un terme générique englobant l'apprentissage, l'éducation et la formation.
[10] Malheureusement, c'est la quasi totalité des pays du Sud qui n'ont ni la capacité économique, ni suffisamment de ressources humaines pour participer.

Le processus normatif instaure au niveau des institutions nationales de normalisation (l'AFNOR en France), mais aussi au niveau de l'assemblée quasi-mondiale des États (ISO), un cadre de gouvernance et de saine émulation des produits et services qui, sans cela, livrés à leur seule concurrence, paralyseraient tout développement tant soit peu harmonieux des technologies. Cela est particulièrement vrai dans les TIC. La loi de Moore[11] par exemple est une pure conséquence de la normalisation convergente et interopérable des composants, des logiciels et des réseaux : les industriels assurés par les normes de trouver un marché pérenne, conséquent donc rentable peuvent investir tout en restant concurrents sur la production mais non sur la différentiation stérile de standards non compatibles.

Reste à savoir qui participe à ces instances, qui en a la possibilité tant en ressources humaines que financières ? L'ISO est ainsi organisé comme une instance parente de l'ONU selon le principe « un État, une voix » (Toussaint, 2003). Ce principe favorise les États les plus riches qui, au SC36 sont presque exclusivement situés au Nord. Parmi les 28 états membres actuels, la force décisionnelle, bien que théoriquement remise à un consensus universel, reste largement marquée par les couleurs des pays du Nord et tout particulièrement des pères fondateurs partagés entre francophones (France, Canada) et anglo-saxons (Etats Unis, Royaume Uni et Australie). Notons aussi leur diversité linguistique : un minimum de 19 langues et de 7 écritures[12].

Le développement d'un cadre normatif des TICE respectueux des spécificités culturelles, linguistiques, économiques, géopolitiques reste cependant possible. Le Sud-est asiatique qui est confronté à une forte disparité culturelle et linguistique par rapport au pôle dominant nord-américain est de ce point de vue un allié précieux à la Francophonie (Hudrisier, 2006, a). En effet, en l'absence de contributions donnant une visibilité numérique à ces disparités dans des ressources numériques *ad hoc*, l'e-Learning du futur pourrait se développer comme une stricte annexe de l'e-Learning nord américain et anglophone. Nous reproduirions dès lors pour l'e-Learning l'équivalent de l'"hyperculture" globalisante d'Hollywood (Tardy &

---

[11] formulée pour la première fois par Gordon E. Moore en 1965, elle postule le doublement annuel des performances des circuits intégrés (mémoires et processeurs). Moore a revu son estimation en 1975 : doublement tous les 18 mois. Il estime qu'elle se poursuivra jusqu'en 2017, date à laquelle elle devrait rencontrer des contraintes physiques. Dares Vole (Michel) in Wikipédia .

[12] En fait beaucoup plus si on compte toutes les langues associées aux grandes langues nationales ce qui augmente encore le nombre d'écritures (7 écritures sur un billet de banque chinois)

Farchy, 2006) et nous devrions en subir bien sûr les conséquences de domination économique mais surtout universitaire et scientifique qui en découleraient.

**2.1 Les effets techno-économiques de la normalisation des TIC**

La normalisation nationale et internationale a été amorcée dès la fin du 19$^{ème}$ siècle pour deux causes convergentes :

- Permettre aux Etat nationaux, ou déjà à l'Europe et l'Amérique industrielle, de contrôler la sécurité et les désirs d'interopérabilité des professionnels et des usagers d'une activité comme par exemple dans le cadre de la Commission Electronique Internationale (CEI), fondée en 1906 qui définit des niveaux normalisés correspondants à des caractéristiques de courants électriques pour en sécuriser les modes de distribution. Un autre exemple est celui de l'Union Internationale des Télécommunications (UIT) qui partage les longueurs d'onde radio pour en assurer une attribution rationnelle, normalisée et stable utile tant pour les industriels que pour les diffuseurs et les utilisateurs.

- Parallèlement, pour les acteurs industriels, mais aussi pour les chercheurs en industrie, s'instaure un cadre normatif suffisamment précis, mais cependant ouvert pour que puisse se développer des produits (composants) normalisés semi-finis (comme en métallurgie, en chimie ou dans l'industrie alimentaire).

Cette deuxième propriété de la normalisation est très importante. Cela permet d'ailleurs de distinguer nettement normes de standards ; et cela permet de comprendre la relation dialectique entre les deux. Le développement industriel moderne exige que se développent non pas seulement des standards, (des innovations isolées, des produits ou services industriels que pourrait maîtriser un seul entrepreneur), mais aussi des *familles de normes* qui s'avèrent indispensables pour déployer comme en e-Learning des réseaux, des ressources partagées, des plates-formes pédagogiques susceptible de s'adapter à des situations linguistiques, institutionnelles, géographiques, disciplinaires extraordinairement diversifiées. Les industriels des composants électroniques avaient déjà compris cela depuis très longtemps. Dans l'après 1$^{ère}$ Guerre mondiale, la radio s'était développée comme un média de masse parce qu'elle avait été rendue disponible à bas coût, grâce à un composant normé, fiable, sécurisé et fabriqué en très grande quantité : la lampe radio ou le TM (tube électronique militaire). C'est selon la même logique que ces mêmes industriels de l'électronique peuvent produire à prix constant des ordinateurs de plus en plus puissants et miniaturisés.

Comme chaque acteur industriel (ou offreur de services) est capable de fabriquer en série (selon des standards industriels souvent gardés secrets) tel ou tel composant des grands ensembles techniques définis par la normalisation des TIC, dès lors nous assistons à un co-développement concurrentiel auto-défini en consensus, puis autogouverné par les acteurs d'un domaine d'activité.

**2.2 Le fonctionnement de la gouvernance normative**

Ce retour aux fondements de la normalisation comme l'une des composantes primordiales de l'expansion industrielle était indispensable pour mieux comprendre comment s'est fondé puis aménagé le système institutionnel mondial de la normalisation. Selon des modes de gouvernement directement inspirés de Saint-Simon, la gouvernance nationale, puis par effet de rebond, la gouvernance mondiale des normes, prévoit une synergie rationnelle de trois catégories d'acteurs participants : les industriels, la puissance régalienne et les usagers. Cette gouvernance est aussi fondée sur une primauté de la raison comme dans les projets d'organisation du pouvoir chez Saint-Simon (Garnier, 1850). La gouvernance des normes n'accepte pas pour verdict définitif le vote de la majorité. Tout projet de norme doit s'organiser en amont pour privilégier un maximum de consensus préalables aux étapes de vote. Tout animateur d'un groupe ou d'un sous-groupe de normalisation sera prioritairement évalué sur sa capacité à animer de façon consensuelle le développement des projets de normes puis leur rédaction. Autre garde fou privilégiant les consensus, tout vote négatif devra être obligatoirement et rationnellement argumenté[13]. Ainsi, d'étape de vote en étape de vote successif, on améliorera les versions d'un "*process*" de production de norme[14] en intégrant, en réfutant ou en amendant ces commentaires à chaque stade intermédiaire.

Cette gouvernance de l'organisation mondiale de la production des produits et services pourrait en première analyse apparaître comme un handicap à la liberté d'entreprendre des industriels (Hudrisier, 2006, a). On a vu qu'il n'en était rien et que la majorité d'entre eux, bien au-delà de s'en accommoder, voyaient là une dynamique, une composante essentielle de leur démarche marketing ou de leurs études de clients-usager. Encore faut-il tempérer cette affirmation en signalant qu'il n'y a pas que les États les plus pauvres qui sont exclus des débats normatifs. Pour les entreprises, entretenir des experts dans ces comités coûte cher, tant pour la délégation de personnel et l'analyse des informations acquises, que par le coût des missions internationales. Là encore, les avantages concurrentiels que procure la participation comme expert à l'élaboration des normes profitent particulièrement aux acteurs industriels les plus riches.

---

[13] Les votes positifs peuvent aussi bien sûr être assortis de commentaires.

[14] Quand les discussions des experts sont globalement terminées, commencent les phases de rédaction provisoire du *Comeettee Draft*, (CD), puis celle du *Final Comeettee Draft*, (FCD), puis le Draft de norme expérimentale (*Draft International Standard*, DIS), puis de projet final de Norme internationale (*Final Draft international Standard*, FDIS) qui précède la mise en place au bout de 3 ans d'expérimentation d'une Norme Internationale. Celle-ci sera révisable tous les 5 ans.

On peut estimer (vu les délais qu'impose la chaîne de *«process»* signalée en note ci-dessus), que des experts impliqués dans des comités de normes connaissent (et même, bien sûr, influencent) les grandes lignes d'une norme avec 3 ou 4 ans d'avance. C'est donc un avantage concurrentiel indiscutable qui profite aux industriels les plus riches. D'autre part c'est pour eux un mode de veille technologique prospective indépassable, car aucun veilleur non impliqué (si habile soit-il) ne pourra se procurer les documents de travail des experts qui sont par nature confidentiels.

C'est d'ailleurs sur cette capacité d'avantage prospectif et de capacité d'influence que s'argumente (entre autres raisons), la légitimité de la participation de l'Agence Universitaire de la Francophonie (AUF) au SC36.

On doit d'ailleurs noter que l'UIT (Union Internationale des Télécommunications) est, aujourd'hui encore, organisée de façon notablement différente quand au débat de gouvernance normative. Pour des raisons stratégiques (notamment militaire), les Etats s'étaient historiquement attribué une place de dominance quasi absolue dans la définition normative des télécommunications. Le partage non seulement des fréquences mais aussi des places nécessairement limitées sur les orbites géostationnaires a conduit les Etats à négocier eux même par le truchement des grandes agences de télécommunications (au début presque toutes nationales). Les industriels (souvent soumis eux-mêmes aux contraintes d'une « politique industrielle d'arsenal ») n'ayant dans ce dialogue que des voix consultatives. La déréglementation et la privatisation massive des télécoms, auxquelles s'est ajoutée l'irruption de l'Internet et de ses pratiques propres de dialogue para normatif avec le W3C ont notablement transformé le paysage. Contrairement à ce que beaucoup de gens croient, le SMSI n'était pas seulement un Sommet mondial organisé généreusement sous l'initiative des Etats pour « comprendre la société mondiale de l'information ». Ce fut avant tout une initiative de l'UIT pressée de rattraper son déficit de gouvernance équilibrée entre les 3 composantes classiques : la puissance régalienne, les industriels associés à la recherche et les utilisateurs. C'est maintenant chose faite après Genève, Tunis et les différents comités de suivis qui ont été organisés.

Pour autant l'équilibre de gouvernance normative mondiale des TICE introduit des enjeux sociétaux qui vont bien au-delà des seuls risques économiques. Certes, la plupart des chercheurs et des industriels[15] des TICE sont engagés dans cette gouvernance planifiée du développement du système technologique mondialisé, mais quelle place y est réservée aux Etats les plus pauvres ou les moins informés [16] ? Quel Etat serait en situation non seulement de défendre mais d'aménager par ses contributions des propositions normatives préservant dans le futur plan d'urbanisme des TICE normalisées un espace pour les diversités du Sud ? Ce vide de gouvernance respectueux des différences, les Etats ou les grandes multinationales, pris individuellement, sont le plus souvent incapables d'y contribuer ou même de le défendre. Il leur apparaît donc non seulement juste mais aussi économiquement opportun de voir ce rôle assuré par l'AUF.

---

[15] On peut repérer quelquefois certains acteurs (par exemple Microsoft), qui semblent maîtriser des pans entiers du potentiel de développement du système technologique mondial. Profitant de l'avantage de leurs standards omniprésent ils cherchent alors à court-circuiter les procédures d'adoption des normes pour imposer leurs standards comme normes. La situation plus classique consiste en ce que les grands acteurs industriels s'entendent en consortium pour tirer aussi (ce qui est bien normal) avantage de leurs recherche-développement.

**3. L'AUF, une fonction relai pour la diversité au sein du SC36**

De ce point de vue, et pour le cas ici traité des normes de l'e-Learning, l'AUF, par sa Liaison[17] catégorie A à l'ISO depuis 2000, est donc en situation de porter le débat sur les questions éthiques, économiques et culturelles qu'un club de pays presque exclusivement situés au Nord a du mal à formuler seul. L'AUF répond ainsi à l'une de ses missions essentielles, celle d'un vecteur de médiation des enjeux stratégiques et des décisions internationales vers ses partenaires du Sud et d'un relai pour porter leurs voix dans les sphères d'influence qui leurs restent encore inaccessibles par manque de moyens ou de volonté (Ben Henda, 2006). Ainsi, elle a été opératrice de deux Open forums des normes de l'e-Learning, l'un à Versailles en 2003 et l'autre à Tunis en novembre 2005 en marge de la tenue de la deuxième session du SMSI. En coopération avec le Sénégal, elle sera l'hôte de la première Plénière africaine de l'ISO-SC36 à Dakar en septembre 2009, démultipliant ainsi les occasions d'ouvertures sur un débat plus vaste autour des questions de la diversité en e-Learning.

L'AUF dispose d'un potentiel de négociation important au sein du SC36. Même si elle n'est pas autorisée à voter les résolutions finales des normes, elle dispose d'une voix délibérative dans les phases préalables de réalisation technique des consensus. Sur ce plan, l'AUF a l'avantage de défendre les intérêts d'une mosaïque de contraintes diversifiées : l'éducation en français mais aussi dans un grand nombre de langues partenaires, selon des contextes institutionnels économiques et géographiques représentatifs du monde entier (plus de 70 pays). Cette situation stratégique de l'AUF qui se positionne de fait comme un réseau d'universités

---

[16] On peut s'étonner que les Emirats du Golfe par exemple ne participent pas pleinement à ces activités normatives hautement stratégiques pour leur développement industriel à moyen et long terme.

[17] Dans le jargon de l'ISO, être « Liaison technique » c'est être lié à une autre instance de normalisation comme par exemple le sous comité technique (TC37) de Terminologie à l'ISO. Etre « Liaison catégorie A ou C » c'est être lié à une institution autre en dehors du cadre de l'ISO comme par exemple l'AUF. L'ensemble des délégations, c'est-à-dire les NBs & les Liaisons, est désigné sous le sigle NBLO (National Bodies and Liaison Organisations)

confrontées obligatoirement à la disparité de leurs institutions nationales et diversité des langues partenaires, s'avère très précieuse pour la quasi-totalité des experts de ces universités. Dans leurs délégations nationales respectives, ils doivent défendre des objectifs nécessairement beaucoup plus étroits : un potentiel national de l'industrie des TICE ainsi que leur mise en oeuvre sous forme d'une offre de ressources, un système éducatif globalement homogène dont il faut défendre et préserver les spécificités, préserver aussi le potentiel d'enseigner dans une ou quelquefois plusieurs langues nationales. Notons que sur ce point chez les experts réellement présents à l'international, il n'est que les canadiens pour affirmer haut et fort la défense d'une culture académique obligatoirement bilingue. Sauf que l'objectif de l'AUF est plutôt d'aller au-delà d'un bilinguisme ou d'un multilinguisme dominant qui concernent des langues « fortes » comme l'anglais et le français ou qui caractérisent des langues largement médiatisées comme les langues européennes. Son souci dans le partenariat francophone est d'impliquer autant que possible des langues et des cultures moins connues et moins représentées dans le débat international en cours autour de la diversité culturelle et linguistique.

De ce fait l'AUF dispose d'une liberté et d'une ouverture d'expertise unique dans le collège des experts du SC36. Cela lui confère très souvent le rôle d'arbitrer certaines controverses dans des situations d'affrontements qui paralysent trop souvent les débats. Citons comme exemple à cette situation d'arbitrage, le cas des pays du Sud-est asiatique qui ont toujours été très enclins à exprimer, à travers les initiatives de l'AUF, l'affirmation de leur volonté de prendre désormais toute la place qui leur revient dans le débat de gouvernance normative des TICE. Ils entreprennent particulièrement la défense de leur industrie des TIC et des TICE extraordinairement prospère et innovante grâce à la culture industrielle héritière du mode de production asiatique. Ils s'activent également dans la défense d'une culture éducative globalement homogène sur le sous-continent, qui présente pour caractéristique principale une disparité très forte avec l'Occident héritier, quant à lui, d'une tradition grecque de la logique et du raisonnement. Leur arrimage du fait éducatif sur l'individu, la famille[18] et la société est lui aussi très éloigné de l'Occident. Tout ceci réfère, bien que sur des modes plus diversifiés, à la disparité des langues du Sud-est asiatique et leur influence sur la totalité de la culture de transmission du savoir.

Cette alliance objective de la diversité culturelle et linguistique du Sud-est asiatique avec l'AUF explique notamment que les deux premières langues développées dans le projet Cartago (voir ci-dessous), aient été le coréen et l'arabe.

---

[18] Voire l'entreprise pour ce qui est de la formation.

On peut aujourd'hui constater, que dans les délégations ayant le statut de liaisons de rang A auprès du SC36, à côté des structures prestigieuses comme l'IEEE[19] et ADL[20], l'AUF gagne en visibilité évidente grâce à ses apports sur la diversité linguistique et culturelle au profit du SC36. Son portefeuille de langues, de cultures et d'États non représentés abonde dans le sens des partisans de la liberté et de la diversité : soit pour des raisons de scrupules éthiques, globalement partagés par nombre d'experts en éducation, soit aussi, sans exclusive de la cause précédente parce que la diversité ouvre de larges perspectives de marchés dans lesquels les nouveaux entrants industriels peuvent plus facilement s'opposer aux opérateurs historiques.

**4. La recherche-action francophone dans la normalisation e-Learning**

Pour s'acquitter de cette tache historique de rôle relai ou d'intermédiation au profit d'un large éventail de structures partenaires dans l'enseignement et la recherche[21] mais qui restent encore absents de ces cercles de décisions stratégiques, l'AUF déploie une série de programmes de Recherche & Développement (R&D) et de stratégies d'action, destinés à renforcer ce créneau technologique porteur de la normalisation en e-Learning vers un auditoire francophone et de langues partenaires au Sud (Ben Henda et al., 2007). Les objectifs sont stratifiés selon un échelonnage de sensibilisation des masses, d'intégration des acteurs en éducation et de production de services et ressources pour une participation active dans les réseaux de l'e-Learning mondial. Potentiellement la quantité de pistes ouvertes par ces programmes est considérable et dépasse bien entendu la capacité de travail et de financement des chercheurs et experts mobilisés sur ce chantier de normes e-Learning. Mais, des efforts concluants ont permis jusqu'ici d'identifier des axes de travail susceptibles de faire évoluer la normalisation de l'e-Learning pour qu'elle soit dans l'avenir un cadre plus équitable, plus respectueux des diversités notamment culturelles et linguistiques.

**4.1 L'appropriation du référentiel scientifique et technique spécialisée de l'e-Learning**

La normalisation de l'e-Learning touche beaucoup des domaines connexes comme les TIC et les systèmes d'information. Elle implique également beaucoup d'aspects inhérents aux sciences de l'éducation comme la pédagogie d'apprentissage et les profils des apprenants, des sciences de l'information et de la communication comme les schémas de métadonnées pour l'indexation des ressources et la description des offres de formation. Le suivi concret des travaux du SC36 et d'un certain nombre d'autres instances normatives ou de standards directement liés aux TICE, est pour le moment relativement sous-exploité alors qu'il représente un énorme potentiel de transfert de technologie. Nous avons bon espoir de prévoir qu'une diffusion massive des documents techniques officiels autour de ces questions, toutefois traduits dans les langues partenaires, aiderait à renforcer une culture technologique locale autour de la normalisation e-Learning. L'objectif est de parvenir à organiser avec des universités du Sud francophone un réseau de traduction et d'exploitation documentaire des ressources recueillies.

---

[19] IEEE : Institute of Electrical and Electronics Engineers. [http://www.ieee.org, visité le 4 avril 2007]

[20] ADL : Advanced Distributed Learning [http://www.adlnet.org, visité le 4 avril 2007]

[21] L'AUF est un réseau de plus de 600 structures d'enseignement et de recherche ayant le français comme l'une des langues de travail.

Cette idée de traduction d'un référentiel documentaire autour de la normalisation des TICE et de l'e-Learning prend racine dans un constat de fait sur l'état de l'art de la combinaison entre les langues courantes véhiculaires d'une information scientifique et technique et les champs thématiques couverts par les ressources documentaires appropriées. Agissant dans un contexte francophone, ou d'une ou plusieurs langues partenaires, les francophones sont généralement confrontés à une documentation majoritairement anglo-saxonne quand il s'agit des TICE et de la normalisation en e-Learning. Historiquement maîtres d'ouvrage en la matière, des structures américaines pionnières de normalisation de TICE et d'e-Learning comme AICC, IMS, IEEE, et tant d'autres, ont marqué de leur culture pédagogique (unilinguisme anglophone, pédagogie procédurale, primauté de la formation sur l'éducation…) les grandes lignes du projet du SC36 en confortant l'idée chez les industriels de ce marché pionnier, que cela était reproductible quasi à l'identique sur le marché de l'e-Learning dans sa phase de mondialisation et de banalisation des usages. Bien que la langue française parvient aujourd'hui à pallier cet écart linguistique et culturel en traduisant la majeure partie du référentiel technique des normes et standards, quand il n'est pas produit à la source en français, il n'en est pas de même pour ses langues partenaires qui restent tributaires d'un relai linguistique francophone pour s'approprier des nouveautés technologiques et normatives dans ce domaine. La traduction de ce référentiel scientifique vers les langues partenaires de la francophonie devient indispensable pour bâtir les ponts nécessaires au transfert de technologie et répondre à la question stratégique du droit à l'éducation pour tous dans la conformité avec les normes et standards internationaux en vigueur.

**4.2 La formation des formateurs**

La formation de formateurs ou de décideurs des universités aux normes et standards de l'e-Learning est un point sur lequel on agit le plus souvent possible soit directement (colloques, sessions de formations, publications), soit plus individuellement par l'implication de nombre d'étudiants de 3$^{ème}$ cycle.

L'action phare de l'AUF sur ce point est sans doute son engagement sur un plan d'action-recherche et de co-développements en matière de technologies appliquées à l'enseignement et à la formation des formateurs via son programme « Soutien des TIC au développement de l'enseignement supérieur et de la recherche », dont le mobile est de mettre les forces productives des pays francophones du Sud au cœur même de la stratégie d'action autour des TICE. L'objectif en est de « *renforcer les capacités humaines par la formation, de réduire la fracture numérique en accroissant la connectivité des universités du Sud, de développer une politique de contenus scientifiques francophones, de favoriser la recherche (réseaux de recherche, observatoire), et la présence francophone dans les comités internationaux (standardisation, normalisation, régulation)* »[22].

Le programme Transfer de formation en ligne, constitue sans doute l'un des acquis clés de cette stratégie d'action envers les partenaires francophones du Sud dans la mesure où il constitue le levier par excellence pour offrir des voies grandes ouvertes vers le savoir faire technologique mondial en éducation. Trois ateliers parmi dix sont réservés aux techniques et standards e-Learning. Le dernier en date (2007), défini autour de la conception standardisée des contenus d'apprentissage, a anticipé le rebondissement retentissant de la soumission récente au domaine public (avril 2007), et donc à la gouvernance mondiale, du standard Scorm, l'un des grands référents dans le monde de l'e-Learning. Devenu membre liaison de catégorie A auprès du SC/36 depuis 2005, ADL, régent des spécifications Scorm au nom du ministère de la défense américaine (DoD) et de la Maison Blanche, lève son monopole sur un référent légendaire dans le monde de l'e-Learning au profit d'une gouvernance mondiale. Gain de cause pour la communauté internationale où repositionnement stratégique intelligent de la part des acteurs influents de la standardisation de l'e-Learning mondial ? Les conséquences sont toutefois probantes : à part le passage au monde du libre et de l'Open source de l'un des grands baliseurs mondiaux de la standardisation de l'e-Learning, la fourniture d'une solution libre de droit pour les communautés des pays émergents leur permettra de s'inscrire dans les systèmes d'accès, d'échange et de développement des ressources pédagogiques qualifiés RAID (réutilisables, adaptables, interopérables et durables). Une gouvernance mondiale de Scorm ouvrirait largement les voies de son internationalisation pour répondre à toutes les spécifications linguistiques et culturelles des acteurs concernés par l'éducation et la formation professionnelle à distance. Les conséquences de l'implication de l'AUF dans ce créneau sont immédiates, dans la mesure où des initiatives de localisation des systèmes auteurs d'e-Learning conformes à Scorm et à ses soubassements référentiels IMS sont déjà en cours, sous une impulsion francophone d'appui aux langues partenaires. Sur un plan prospectif, la touche multilingue et multiculturelle apportée à cette thématique de formation aurait comme conséquence le renforcement d'une caractéristique clé des normes et des standards : l'interopérabilité. Plus l'interopérabilité entre des dispositifs d'e-Learning est fiable, plus les chances de mutualisation et d'échange des ressources pédagogiques sont grandes entre les acteurs et les partenaires de l'e-Learning public et privé. La diversité linguistique et culturelle venant imprégner ces dispositifs, il en résulte une accentuation de la production des ressources éducatives locales et par conséquent, la création de réservoirs pédagogiques nationaux.

---

[22] AUF. Soutien des TICs au développement de l'enseignement supérieur et de la recherche. [http://www.auf.org/rubrique21.html, visité le 5 avril 2007]

**4.3 L'alliance Cartago, un levier pour la terminologie multilingue normalisée**

Dans ce chantier normatif e-Learning, l'une des préoccupations majeures de l'AUF, associée à la diversité culturelle et linguistique, est sans doute cette évolution vers la nouvelle génération des TICE fondée sur le langage XML (eXtensible Markup Language) qui ne cesse de susciter un certain nombre de développements et d'expérimentations tant pour des collèges virtuels d'étude de corpus que pour des propositions de pédagogie fondées sur une collégialité sémantique. Du point de vue des instances de normalisation (ISO, W3C…), ces approches n'ont un avenir qu'autant que les différents acteurs de ces nouveaux usages des TICE reprendront à leur compte la nécessité de s'inscrire dans des instances normatives leur permettant de construire les terminologies et les ontologies normalisées représentant les fondements conceptuels et épistémologiques partagés de leur discipline. C'est en quelque sorte construire les prémices de ce qui est communément appelé le Web sémantique. Sauf que, s'il est toujours vrai que le Web sémantique offre un cadre d'intelligence additive pour l'exploitation des ressources d'information, ceci n'est rendu possible que si les domaines de connaissance aient été déjà décrits par des acteurs humains dans leurs langues et cultures respectives, qu'ils soient des spécialistes du domaine lors de la constitution d'ontologies de référence ou des auteurs des documents lors de leurs activités d'annotation à l'aide de métadonnées.

Ceci remet en évidence toute l'importance octroyée aux chapitres du vocabulaire et de la terminologie, développés depuis 1979 par Eugen Wüster dans sa Théorie Générale de la Terminologie (Wüster, 1979). Cette théorie est fondée sur un ensemble de mécanismes fondamentaux pour construire des systèmes de gestion de corpus lexicaux homogènes, partageables et interopérables dont l'objectif est de concevoir et mettre en place des bases de données terminologiques multilingues capables d'alimenter les schémas de métadonnées en ressources lexicales et de construire des ontologies spécialisées dans toutes les langues et pour toutes les cultures. Le SC36 consacre deux de ses groupes de travail (parmi sept), pour se concentrer sur ces questions : le groupe de travail 1 (WG1) chargé de la terminologie e-Learning et le groupe de travail 7 (WG7) concentré sur les questions de l'adaptabilité culturelle et linguistique et de l'accessibilité dans les normes en devenir. L'AUF joue un rôle clé dans ces deux structures pour élargir tant que possible la couverture des langues et des cultures des entités non représentées dans le SC36. Le projet Cartago, décrit ci-après, en est l'instrument et le cadre dans lequel ces questions sont formellement soulevées et défendues par la Francophonie dans la rigueur des débats autour des normes.

Le projet et l'Alliance Cartago ont été mis en place à Tunis lors du SMSI 2005 entre une douzaine d'experts du domaine (SC36, AUF, Union Latine) répartis sur tous les continents (Asie, Australie, Afrique, Amérique, Europe). Cartago est une banque terminologique normalisée en phase de développement (Hudrisier, 2005). Elle est dédiée à l'e-Learning dans un nombre extensif de langues. A ce jour, l'arabe, l'anglais, le berbère, le coréen, le français, le malgache, sont déjà disponibles. Le chinois, l'hébreu, le roumain, le vietnamien sont en cours de développement et nombre d'autres langues devraient venir rejoindre le projet. Cette base terminologique est onomasiologique. On y part du concept pour pointer vers des termes en n langues ; ce qui permet une meilleure articulation d'interopérabilité et de synergie entre les cultures et les langues que la démarche sémasiologique de direction inverse. Cette banque construite en XML fondée sur Génétrix (un cadre de données terminologique adopté par l'ISO) dont le créateur André Le Meur (Université de Rennes) collabore directement dans l'équipe de la liaison AUF auprès du SC36. Cartago devient intéressant du moment qu'il permet de servir de référentiel sémantique commun en laissant à chaque communauté linguistique sa capacité équitable à exprimer ses particularismes conceptuels : liés notamment à la capacité de transmettre des savoirs et à rendre intelligibles les modes d'organisations sociétaux propres à telle ou telle communauté.

La plus value de cette démarche de recherche-action est la genèse d'une dynamique de groupe qui rallie des partenaires internationaux de la francophonie autour d'une procédure de travail commune, basée sur l'application de standards et de normes internationaux avec, comme conséquence directe, la transposition des acquis de cette initiative sur les contextes nationaux de l'e-Learning. Le projet Cartago, fondé à l'origine sur la reprise et la duplication multilingue de la terminologie e-Learning du SC36, référencée dans le document ISO 2382-36, procède à l'élaboration des équivalents linguistiques de la terminologie e-Learning, sur la base d'une validation officielle de la part des instances nationales compétentes pour chaque langue ou communauté. Ce fut, entre autres, le cas de l'IRCAM[23] pour le berbère et du Bureau de la Coordination de l'Arabisation pour les pays arabes (organe de l'Alecso[24]) qui ont vu, sous l'impulsion de coordonnateur de projets Cartago, leurs bases terminologiques actualisées voire enrichies par des termes et concepts nouveaux. L'e-Learning, n'étant pas encore dans les habitudes et les pratiques communes dans beaucoup de pays du monde, la terminologie qui lui est associée reste encore largement utilisée en anglais et en français. La reproduction multilingue de la terminologie e-Learning s'inspire largement des définitions dans l'une de ces deux langues. Ainsi, l'initiative francophone aurait accompli à travers le projet Cartago, l'un de ses rôles médiateurs envers ses langues partenaires dans le domaine de l'éducation[25]. Nous considérons, que cette dimension, aussi granulaire soit-elle dans le processus e-Learning, pourrait être d'un appui considérable aux politiques nationales de l'enseignement à distance encore en gestation dans beaucoup de pays émergeants. Beaucoup de ces pays sont encore en phase de structurer leurs cadres globaux d'e-Learning comme le CVM au Maroc (Campus Virtuel Marocain) ou l'UVT en Tunisie (Université Virtuelle Tunisienne). Certains ont déjà progressé vers le stade de la conception des contenus d'apprentissage. Une fois la phase du référencement des ressources pédagogiques sera atteinte, le besoin de recourir à des métadonnées standardisées sera ressenti et le recours à une terminologie multilingue sera une nécessité pour des pays et des cultures bi ou multilingues. Le précédent français du projet Lom.fr[26], mis en place pour indexer le référentiel pédagogique des universités françaises traduit bien cette anticipation sur une progression vers un besoin de terminologie standardisée.

---

[23] Institut Royal de la Culture Amazighe au Maroc qui s'est coordonné en consensus avec les travaux équivalents au « Commissariat algérien » et au CNED (mission berbère au Baccalauréat.)
[24] Alecso : L'Organisation Arabe pour l'Education, la Culture et les Sciences. http://www.alecso.org.tn [Visité le 5 avril 2007]

**4.4 Le besoin d'une gouvernance e-Learning mondiale**

La mondialisation numérique de la transmission du savoir notamment *via* l'e-Learning est désormais un fait. Cette activité humaine connaît une dynamique d'expansion, d'innovation de convergence avec les autres secteurs médiatiques créant « *des sources d'innovation organisationnelles et commerciales substantielles, voire des vecteurs de création d'offres originales et d'exploitation de nouvelles opportunités de marché* » (Charlier et al., 2006). Selon Guillemet, de télé-université, Canada, cette rencontre entre milieux éducatifs et industriels n'est point nouvelle. Elle est par contre marquée cette fois par « *l'intensité du volontarisme, l'ampleur des enjeux institutionnels, l'importance des ressources techniques, humaines et financières, et surtout la convergence des initiatives et l'étroitesse de la coordination* [qui] *semblent créer une situation nouvelle* » d'un fonctionnement des sphères éducatives selon des mécanismes désormais moins artisanaux, certes conformes à des logiques de marché, mais sans pour autant se concentrer exclusivement sur la quête de profits (Guillemet, 2004).

Il est d'autant plus heureux que ce secteur d'activité ai été proposé comme devant être soumis à normalisation internationale parce que s'enclenche dès lors un cercle vertueux de programmation prospective du développement et de la recherche des TICE, qui permet, si la plupart des acteurs jouent « suffisamment honnêtement le jeu », de dépasser les affrontements souvent stériles entre grands acteurs industriels cherchant à imposer leur standards comme monopole mondial. Si tout se passe avec une transparence suffisante, le SC36 devrait être à même de construire le cadre normatif permettant une convergence et une interopérabilité des TICE successibles de répondre aux enjeux d'éducation et de formation dans le monde. Mais, il faut avoir suffisamment de pragmatisme pour ne pas être angélique, afin de reconnaître l'intérêt du secret et de la dissimulation comme dynamique indispensable de l'innovation industrielle et de la santé du marché. On oserait le néologisme de macro transparence nécessaire et suffisante : ne pas avoir le droit de mentir quand les grands enjeux éducationnels mondiaux sont en jeu. Répondre à cet enjeu notamment dans les pays du Nord mais aussi chez un nombre extensif de ressortissants des pays du Sud est véritablement un enjeu vital. Notons, toutefois, que les produits d'e-Learning répondant à ces besoins impérieux de démultiplication pédagogique ne peuvent pas seulement être diffusés dans les grandes langues des pays développés. Pour être efficaces ils doivent pouvoir être conçus et doivent être opérationnels dans toutes les langues du monde y compris, notamment dans des langues africaines, pacifiques, amérindiennes, par voie écrite, par voie orale et audiovisuelle impliquant toutes les formes innovantes de l'éducation en ligne tels le e-Learning, le m-Learning, le t-Learning ...

---

[25] Cartago prétend réparer des « pertes de domaines linguistiques dans des langues dominées (concept qui a donné son nom à un département de l'Université de Paris 8), et qui prétend aussi à participer comme frein à la « fracture numérique ».

[26] Lom.fr : profil français d'application du Learning Object Metadata (LOM) par le Ministère de l'Education français. http://www.educnet.education.fr/articles/lom-fr.htm [Visité le 5 avril 2007].

L'ouverture par la délégation AUF auprès du SC36 d'un chantier de discussion au sein des normalisateurs de l'e-Learning sur la diversité culturelle et linguistique en éducation a constitué un démarrage positif d'une concertation mondiale autour de la gouvernance future de l'e-Learning. L'objectif de l'AUF, mais aussi d'autres instances comme l'Union Latine, le Forum mondial des cultures, l'Asia E-Learning Network, impliquées dans ce débat, serait d'aboutir le plus rapidement possible à une Agence de gouvernance mondiale entre grands acteurs de l'e-Learning pour contrebalancer la seule légitimité des États souverains qui ont du mal à transcender les objectifs mondiaux ; sauf pour les plus puissants qui les accaparent à leur seul profit.

L'hypothèse que l'on pourrait avancer à ce stade s'appuie sur la situation singulière de l'AUF dans le collège des « liaisons du SC36 » où elle représente un avis et une expertise utile à l'équilibre des consensus. Les contraintes de diversités linguistiques, culturelles, institutionnelles, disciplinaires, géopolitiques, économiques etc. met tout le monde devant l'obligation de penser mondialement le développement de l'e-Learning normalisé et interopérable du futur. C'est dans cette perspective que la contribution de l'AUF aux travaux du SC36 a toujours rencontré une adhésion quasi unanime ou même a suscité certaines thématiques émergeantes qui semblent être des conditions importantes de la prospérité du marché à travers les normes de l'e-Learning. L'objectif est de défendre des normes qui garantissent :

- l'accès et la compatibilité de toutes les langues

- un potentiel d'adaptabilité avec des institutions académiques ou de formations très disparates (liés très souvent aux états et aux cultures)

- une adaptabilité aux styles et cultures pédagogiques les plus divers et les plus composites

- une adaptabilité à toutes disciplines et métiers y compris les moins rentables

- une éthique de l'enseignement, un respect des biens commun du savoir et des libertés individuelles

- la possibilité de délimiter les espaces de la marchandisation du savoir d'avec ceux d'une sanctuarisation des espaces non "*marchandisables"*.

## 5. Conclusion

Les questions posées par la mondialisation mais aussi par les grands enjeux de survie à moyen terme de l'espèce humaine (notamment les enjeux écologiques, les enjeux de maintient de la paix, de maîtrise des débordements migratoires…) convergent toutes, on le sait mieux d'année en année, vers une obligation d'augmenter le niveau d'éducation et de formation sans commune mesure avec le niveau actuel. Cette mutation indispensable du niveau d'instruction des hommes et des femmes (et certainement pas seulement des enfants), sans trop d'augmentation de la proportion de dépenses d'éducation et de formation par rapport aux autres secteurs d'activité passe à l'évidence par une démultiplication des capacités d'enseignement que permet l'e-Learning dans des modèles de dispositifs standardisés, compatibles et interopérables pour garantir la mutualisation et l'échange. Ceci rejoint l'optimisme relatif de Jacques Attali qui témoigne, dans son rapport au ministre de l'éducation français, que « *Si* [la mondialisation] *était appliquée à l'éducation, elle conduirait à la mise en place d'un modèle mondial d'enseignement supérieur standardisé, dans lequel l'Etat s'effacerait et le marché modèlerait les cursus et les carrières* » (Attali, 1998). A l'évidence cette démultiplication doit se négocier dialectiquement avec les professionnels de l'enseignement et de la formation et notamment les grands réseaux mondiaux.

Parvenir à baliser normativement le déploiement futur de l'e-Learning vers ses véritables enjeux indispensables pour affronter le futur est certainement plus complexe que de normaliser ce qui existe déjà ou ce qui est déjà programmé dans le développement-recherche des industriels. Si nous ne parvenons pas à encadrer normativement les besoins des grands réseaux mondiaux d'enseignement, ou des grands théoriciens de la pédagogie[27], ils produiront naturellement des outils et des ressources qui dans quelques années, seront plus difficile encore à normaliser. La mission de l'AUF, comme organe francophone bien placé par ses acquis et réalisations dans la sphère de l'éducation mondiale, est particulièrement d'œuvrer pour renforcer l'équilibre normatif de l'e-Learning entre les spécificités linguistiques et culturelles ainsi que les besoins et les attentes technologiques et pédagogiques de tous les acteurs concernés, particulièrement ceux du Sud, par ce secteur stratégique pour le développement durable.

---

[27] Ces réseaux comme ces théoriciens qui pour la majorité d'entre eux nous ignorent encore aujourd'hui ou ne comprennent pas la nécessité de nos actions.

**Bibliographie**